\documentclass[11pt]{article}
\usepackage[dvips]{graphics}

\title{\Large\bf  Vector Riemann-Hilbert problem with almost periodic and meromorphic 
coefficients and applications}
\author{\bf {\sc By Y.A.\ Antipov}\\ 
Department of Mathematics, Louisiana State University\\
Baton Rouge LA 70803, USA}

\date{}

\newcommand{\supp}{\mathop{\rm supp}\nolimits}

\newcommand{\I}{\mathop{\rm Im}\nolimits}
\newcommand{\R}{
\mathop{\rm Re}\nolimits}
\newcommand{\const}{\mbox{const}}

\newcommand{\Md}{\partial}

\newcommand{\Ga}{\alpha}
\newcommand{\Gb}{\beta}
\newcommand{\Gd}{\delta}

\newcommand{\Gve}{\varepsilon}
\newcommand{\Gf}{\phi}
\newcommand{\Gvf}{\varphi}
\newcommand{\Gg}{\gamma}
\newcommand{\Gc}{\chi}

\newcommand{\Gk}{\kappa}

\newcommand{\Gl}{\lambda}

\newcommand{\Gm}{\mu}

\newcommand{\Gt}{\theta}

\newcommand{\Gr}{\rho}

\newcommand{\Gs}{\sigma}

\newcommand{\Go}{\omega}
\newcommand{\Gx}{\xi}

\newcommand{\GD}{\Delta}
\newcommand{\GF}{\Phi}
\newcommand{\GG}{\Gamma}

\newcommand{\GP}{\Pi}

\newcommand{\GO}{\Omega}

\newcommand{\GY}{\Psi}

\newcommand{\CH}{{\cal H}}

\newcommand{\beq}{\begin{equation}}
\newcommand{\eeq}{\end{equation}}
\newcommand{\barr}{\begin{eqnarray}}
\newcommand{\earr}{\end{eqnarray}}
\newcommand{\beqn}{\begin{equation*}}
\newcommand{\eeqn}{\end{equation*}}
\newcommand{\barrn}{\begin{eqnarray*}}
\newcommand{\earrn}{\end{eqnarray*}}

\newcommand{\fr}{\frac}

\begin{document}
\maketitle

\begin{abstract}

The vector Riemann-Hilbert problem is analyzed when the entries of 
its matrix coefficient are meromorphic and almost periodic functions. 
Three cases for  the meromorphic functions, when  they
have (i) a finite number of poles and zeros  (rational functions),
(ii) periodic poles and zeros, and (iii)  an infinite number of non periodic
 zeros and poles, are considered.
 The first case is illustrated by the heat equation for a composite rod
 with a finite number of discontinuities and a system of convolution 
 equations; both problems are solved explicitly. In the second case, a
 Wiener-Hopf factorization is found in terms of the hypergeometric
 functions, and the exact solution of a mixed boundary value problem 
 for the Laplace equation in wedge is derived. In the last case, the Riemann-Hilbert
 problem reduces to an infinite system of linear algebraic equations
 with the exponential rate of convergence. As an example, the Neumann boundary value problem for the Helmholtz equation
 in a strip with a slit is analyzed.

\end{abstract}

\setcounter{equation}{0}

\section{Introduction}

Many physical models described by boundary value problems for elliptic, hyperbolic,
and parabolic equations reduce to a system of $n$ convolution  
equations on a finite segment. Such systems are equivalent to a vector Riemann-Hilbert problem
(RHP) with a block-triangular matrix coefficient
\beq
G(\Ga)=\left(\begin{array}{cc}
e^{i\Ga} I & 0\\
g(\Ga) & e^{-i\Ga}I \\
\end{array}
\right), \quad \Ga\in (-\infty,+\infty),
\label{1.1}
\eeq
where $I$ is the order $n$ unit matrix and $g(\Ga)$
is an $n\times n$ matrix. The functions $e^{i\Ga}$ and $e^{-i\Ga}$ are almost periodic
functions (Levitan, 1953), and their indices are infinite: ${\rm ind}\, e^{i\Ga}=+\infty$ and
 ${\rm ind}\, e^{-i\Ga}=-\infty$. 
If $n=1$ and $g(\Ga)$  is a rational function, then the RHP admits a closed-form solution (Ganin, 1963).
In the general case of the function  $g(\Ga)$, even when $n=1$, there is no
procedure for solving the RHP with the coefficient (\ref{1.1}) in closed form. Novokshenov (1980)
analyzed the singular convolution equation $\int_0^a[(x-t)^{-1}+k(x-t)]u(t)dt=f(x)$, $0<x<a$, 
$k(x)\in  L_2(-\infty,+\infty)$, and showed that its solvability and the representation
formulas for the solution are obtained in terms of the Wiener-Hopf factors of the associated matrix (\ref{1.1}) ($n=1$). A theory of factorization of matrices (\ref{1.1}) when $g(\Ga)$ is an almost
periodic function (including an almost periodic polynomial) was developed by 
Karlovich and Spitkovsky (1983),  
Spitkovsky (1989)
and their coauthors (see for example 
B\"ottcher et al, 2002).  

In many applications to physical models, $g(\Ga)$ 
is a meromorphic function. Antipov (1987, 1989) considered contact problems on an annular
stamp and reduced it to a RHP with the matrix coefficient (\ref{1.1}),
$g(\Ga)=\fr12\GG(\fr{\Ga}{2})\GG(\Gg+\fr12-\fr{\Ga}{2})[\GG(\fr12+\fr{\Ga}{2})\GG(\Gg+1-\fr{\Ga}{2})]^{-1}$, $\Gg=0,1$. The problem was transformed into an infinite system
of linear algebraic equations of the second kind with the exponential rate of convergence
and solved in terms of recurrent relations. Recently, this technique for employed for the solution of an integro-differential convolution equation arising in fracture with surface effects
(Antipov and Schiavone, 2011).

Another class of matrices of the form
\beq
G(\Ga)=\left(\begin{array}{cc}
g_{11}(\Ga) & e^{i\Ga}g_{12}(\Ga)\\
e^{-i\Ga}g_{21}(\Ga)  & g_{22}(\Ga) \\
\end{array}
\right), \quad \Ga\in (-\infty,+\infty),
\label{1.2}
\eeq
needs to be factorized to solve the systems of convolution equations
$k_{j1}\ast\Gf_1+k_{j2}\ast\Gf_2=f_j(x)$, 
$a_j<x<\infty$, 
$\supp\Gf_j\subset[a_j,\infty)$, 
$j=1,2$,  $a_1=0$, $a_2=a$. A method of integral equations for factorizing
matrices of such a structure was proposed by Abrahams \& Wickham (1990).
Onishchuk (1988), Antipov \& Arutyunyan (1992), Antipov (1995, 2000) worked out a
 technique for vector RHPs when   $g_{ij}(\Ga)$
are  meromorphic functions.
and applied it to static fracture and contact 
problems. This method ultimately requires solving infinite systems
of linear algebraic equations of the second kind with the exponential rate of convergence.

In this paper we aim to develop further the methodology for the RHPs whose matrix  coefficient 
entries are meromorphic and almost-periodic functions and apply it for the solution
of some model physical problems.
In Section 2, we consider the heat equation for an infinite  rod $u_t=a^2(x)u_{xx}+g(x,t)$
with a piece-wise constant diffusivity $a(x)$. Deconinck et al (2014) applied the Fokas method (2008) to find an exact solution in the homogeneous case, $g\equiv 0$, when the diffusivity is
a piece-wise constant function, and  (i)   the rod is finite and the diffusivity
has one or two points of discontinuity, and (ii) the rod is infinite and the function
$a(x)$ is discontinuous either at $0$ and $\infty$, or at two finite points and infinity.
In particular, for an infinite rod with $a(x)=a_\pm$, $\pm x>0$, and $g\equiv 0$, they managed to
represent the solution by double integrals with  one-sided Fourier internal integrals 
and contour external integrals over the boundary of infinite wedge-like domains.
In our case, the rod is infinite, and the diffusivity has $n+1$ points of discontinuity.
We show how the problem can be reduced to an order-$n$ RHP and solved exactly
in terms of the solution of an associated finite system of linear algebraic 
equations. In particular, when $n=1$, and the points of discontinuity are $0$ and $\infty$,
we simplify the solution and derive an exact formula for the temperature. 
If $g\equiv 0$, then the solution is given by a sum of two one-dimensional integrals
over the intervals $(-\infty,0)$ and $(0,\infty)$. The formula found generalizes the classical Poisson formula
for the homogeneous infinite rod. We also briefly describe another approach for the solution
of the heat problem that employs the Laplace transform and the theory of discontinuous one-dimensional boundary value problems. The solution constructed by this alternative approach
coincides with the one found by the RHP technique and is applicable to both cases, finite and infinite, and for any finite number of discontinuities. At the end of Section 2, we present
an alternative approach for solving the Abrahams-Wickham system of convolution equations
(Abrahams \& Wickham, 1990) by reducing it to the case (\ref{1.2}) with $g_{ij}$
 being rational functions. Notice that their procedure for the RHP hinges on
a solution of an auxiliary  system of integral equations, while our approach bypasses extra integral equations.

In Section 3, we propose a factorization method for matrices (\ref{1.2}) when
the functions $g_{ij}$ are meromorphic and their zeros and poles are periodic.
The method is illustrated by solving a mixed boundary value problem for the Laplace
equation in a wedge. This problem for a particular choice of the boundary data
was analyzed by Abrahams \& Wickham (1990). They rewrote the problem as a vector
RHP with the coefficient of the form (\ref{1.2}) and reduced it to a system of two integral equations to  be solved numerically by an approximate method. 
The approach we propose does not require solving any auxiliary
integral equations. It derives Wiener-Hopf factors of the matrix coefficient in terms of the 
hypergeometric functions and ultimately yields a closed-form solution of the physical 
problem.

In Section 4, we generalize the method for RHPs with the matrix coefficient of the form 
(\ref{1.1}) for the dynamic case. As an illustrative example, we take the Nemann boundary value problem for the  Helmholtz equation
in a strip with a finite slit. This problem describes antiplane strain deformation of a strip when the strip boundary is free of traction, and
the Mode-III crack faces are subjected to oscillating loading. At the same time it could be interpreted as a model of sound  transmission in a waveguide when an acoustically hard finite screen is placed
inside the waveguide.  We derive the solution by quadratures
and some exponentially convergent series with the coefficients
determined from a rapidly convergent infinite system of linear equations of the second kind.

\setcounter{equation}{0}

\section{Matrices with almost periodic and rational entries}\label{s2}

In this section we derive an order-$n$ vector RHP associated with the heat equation in a rod
with a piece-wise constant diffusivity and conductivity and show that it admits a closed-form solution.
To verify the procedure, we derive the solution 
by the standard technique of discontinuous one-dimensional boundary value problems that bypasses the RHP.  To give an extra example, we solve 
a system of two convolution equations.

\subsection{Heat equation with piece-wise constant coefficients: RHP for an order-$n$ vector-function}\label{s2.1}

The problem under consideration is one of heat conduction for 
an infinite rod with a piece-wise constant diffusivity $a^2(x)=k(x)/[c_p\Gr(x)]$
$$
u_t=a^2(x)u_{xx}+g(x,t), \quad |x|<\infty, \quad x\ne b_0,b_1,\ldots,b_{n-1}, \quad t>0.
$$
$$
u|_{x=b_j^-}=u|_{x=b_j^+}, \quad k_ju_x|_{x=b_j^-}=k_{j+1}u_x|_{x=b_j^+}, \quad j=0,1,\ldots,n-1, \quad t\ge 0,
$$
\beq
u|_{t=0}=f(x), \quad |x|<\infty,
\label{2.1}
\eeq
where $u(x,t)$ is the temperature, $g(x,t)=(c_p\Gr)^{-1}g_0(x,t)$, $g_0(x,t)$ is the heat source density,
$f(x)$ is an initial temperature, $k(x)$ is the thermal conductivity, $\Gr(x)$ is the density,  $c_p$
is the specific heat capacity, and
\beq
a(x)=a_j>0, \quad k(x)=k_j, \quad \Gr(x)=\Gr_j, \quad x\in(b_{j-1}, b_j), \quad j=0,1,\ldots,n.
\label{2.1'}
\eeq
Here, we assumed $b_{-1}=-\infty$ and $b_n=+\infty$. In what follows we reduce this physical problem to an order-$n$ vector RHP. Introduce first
the Laplace transforms
\beq
\hat u(x;p)=\int_0^\infty u e^{-pt}dt, \quad \hat g(p)=\int_0^\infty g e^{-pt}dt, \quad \R p>0,
\label{2.2}
\eeq
and obtain from (\ref{2.1}) a discontinuous one-dimensional boundary value problem.
It reads
$$
a^2(x) \hat u_{xx}-p\hat u=-f-\hat g, \quad |x|<\infty, \quad x\ne b_0, b_1,\ldots,b_{n-1},
$$
\beq
\hat u|_{x=b_j^-}=\hat u|_{x=b_j^+}, \quad k_j\hat u_x|_{x=b_j^-}=k_{j+1}\hat u_x|_{x=b_j^+}, \quad j=0,1,\ldots,n-1.
\label{2.3}
\eeq
To apply further the two-sided Laplace transform, we 
 introduce new functions of the parameter $p$
$$
\Gb_{0j}(p)=a_j^2\hat u|_{x=b_j^-}-a_{j+1}^2\hat u|_{x=b_j^+},
$$
\beq
\Gb_{1j}(p)=a_j^2\hat u_x|_{x=b_j^-}-a_{j+1}^2\hat u_x|_{x=b_j^+},\quad j=0,1,\ldots,n-1,
\label{2.4}
\eeq
and integrate by parts
\beq
\int_{-\infty}^\infty a^2(x)\hat u_{xx} e^{-sx}dx=\sum_{j=0}^{n-1}(\Gb_{1j}+s\Gb_{0j})e^{-sb_j}+s^2
\int_{-\infty}^\infty a^2(x)\hat ue^{-sx}dx.
\label{2.5}
\eeq
We split now the integral in (\ref{2.5}) into $n+1$ parts and denote
$$
\int_{-\infty}^{b_0} \hat u(x;p)e^{-sx}dx=e^{-sb_0}U_0^+(s),
$$$$
\int_{b_{j-1}}^{b_{j}} \hat u(x;p)e^{-sx}dx=e^{-sb_j}U_j^+(s)=e^{-sb_{j-1}}U_j^-(s),\quad j=1,2,\ldots,n-1,
$$
\beq
\int_{b_{n-1}}^{\infty} \hat u(x;p)e^{-sx}dx=e^{-sb_{n-1}}U_0^-(s),
\label{2.6}
\eeq
where 
$$
U_0^+(s)=\int_{-\infty}^0 \hat u(x+b_0;p)e^{-sx}dx,\quad U_0^-(s)=\int_0^{\infty} 
\hat u(x+b_{n-1};p)e^{-sx}dx,
$$
$$
U_j^+(s)=\int_{b_{j-1}-b_j}^0 \hat u(x+b_j;p)e^{-sx}dx,\quad 
U_j^-(s)=\int_0^{b_{j}-b_{j-1}} \hat u(x+b_{j-1};p)e^{-sx}dx,
$$
\beq
 j=1,2,\ldots,n-1.
\label{2.7}
\eeq
The functions $U_j^+(s)$ and  $U_j^-(s)$ ($j=0,1,\ldots,n-1$) are analytic in the domains 
$D^+=\{ \R s<0\}$ and $D^-=\{\R s>0\}$, respectively. 
We emphasize that except for $U_0^\pm(s)$ all the other functions $U_j^\pm(s)$ 
($j=1,2,\ldots,n-1$) are 
entire functions in the half-planes  $D^\mp$. In these notations, the one-dimensional boundary value problem (\ref{2.3}) can be recast as the following order-$n$ vector RHP:
$$
U_j^+(s)=e^{s(b_j-b_{j-1})} U_j^-(s), \quad j=1,2,\ldots,n-1,
$$
\beq
\sum_{j=0}^{n-1} m_j(s) e^{-sb_j} U_j^+(s)+m_n(s) e^{-sb_{n-1}} U_0^-(s)=H(s), \quad s\in L,
\label{2.8}
\eeq
where $L$ is the positively oriented imaginary axis (the domain $D^+$ is on the left), $m_j(s)=a_j^2 s^2-p$, $j=0,1,\ldots,n$, and $H(s)$ is given by
\beq
H(s)=-\int_{-\infty}^\infty[f(x)+\hat g(x;p)]e^{-sx}dx-\sum_{j=0}^{n-1} e^{-sb_j}(\Gb_{1j}+s\Gb_{0j}).
\label{2.9}
\eeq
The vector RHP (\ref{2.8}) can be transformed into a finite system of linear algebraic equations.
To do this, without loss of generality, we assume $b_0=0$ and rearrange the RHP (\ref{2.8})
as follows:
$$
U_0^+(s)+\fr{m_1}{m_0} U_1^-(s)+\ldots+
e^{-sb_{j-1}}\fr{m_j}{m_0} U_{j}^-(s)+\ldots+
e^{-sb_{n-2}}\fr{m_{n-1}}{m_0} U_{n-1}^-(s)
$$$$
+e^{-sb_{n-1}}\fr{m_n}{m_0} U_{0}^-(s)=\fr{H(s)}{m_0},
$$$$
e^{sb_j}\fr{m_0}{m_j}U_0^+(s)+\ldots+e^{s(b_j-b_{j-1})}\fr{m_{j-1}}{m_j} U_{j-1}^+(s)+U_j^+(s)+
\fr{m_{j+1}}{m_j} U_{j+1}^-(s)
$$$$
+\ldots+e^{s(b_j-b_{n-1})}\fr{m_n}{m_j} U_{0}^-(s)=\fr{e^{sb_j}H(s)}{m_j},\quad j=1,2,\ldots,n-2,
$$
$$
e^{sb_{n-1}}\fr{m_0}{m_{n-1}}U_0^+(s)+\ldots+e^{s(b_{n-1}-b_{j})}\fr{m_{j}}{m_{n-1}} U_{j}^+(s)
+\ldots+e^{s(b_{n-1}-b_{n-2})}\fr{m_{n-2}}{m_{n-1}} U_{n-2}^+(s)
$$
\beq
+ U_{n-1}^+(s)+\fr{m_n}{m_{n-1}} U_{0}^-(s)=\fr{e^{sb_{n-1}}H(s)}{m_{n-1}},\quad s\in L.
\label{2.10}
\eeq
Notice that the functions  $e^{s(b_j-b_l)}U_l^+(s)$ ($l=0,1,\ldots,j-1$) are analytic in $D^+$
and decay exponentially as $s\to\infty$ in $D^+$, while the functions $e^{s(b_j-b_l)}U_{l+1}^-(s)$,
$l=j+1,\ldots,n-1$ ($U_n^-(s)=U_0^-(s)$) are analytic in $D^-$ and decay exponentially
as $s\to\infty$, $s\in D^-$. On factorizing the functions 
\beq
\fr{m_{j+1}(s)}{m_j(s)}=\fr{K_j^+(s)}{K_j^-(s)}, \quad j=0,1,\ldots,n-1,
\label{2.11}
\eeq
where
\beq
K_j^+(s)=\fr{a_{j+1}s-\sqrt{p}}{a_js-\sqrt{p}}, \quad 
K_j^-(s)=\fr{a_js+\sqrt{p}}{a_{j+1}s+\sqrt{p}},\quad \R\sqrt{p}>0,
\label{2.12}
\eeq
and substituting this into equations (\ref{2.10}) we have
$$
\fr{U_j^+(s)}{K_j^+(s)}+\fr{1}{K_j^+(s)}\left[e^{sb_j}\fr{m_0}{m_j}U_0^+(s)+\ldots+e^{s(b_j-b_{j-1})}\fr{m_{j-1}}{m_j} U_{j-1}^+(s)\right]-\CH_j^+(s)
$$
$$
=-\fr{U^-_{j+1}(s)}{K_j^-(s)}-\fr{1}{K_j^+(s)}\left[e^{s(b_j-b_{j+1})}\fr{m_{j+2}}{m_j} U_{j+2}^-(s)+\ldots+e^{s(b_j-b_{n-1})}\fr{m_n}{m_j} U_{0}^-(s)\right]-\CH_j^-(s),
$$
\beq
\quad j=0,1,\ldots,n-1.
\label{2.13}
\eeq
Here,  $\CH_j^+(s)$ and $\CH_j^-(s)$ provide the splitting of the functions
$
e^{sb_j}H(s)[m_j(s)K^+_j(s)]^{-1}
$
 into analytic parts in the domains $D^+$ and $D^-$, respectively.
In general, they are defined by the Sokhotski-Plemelj formulas
\beq
\CH^\pm(s)=\pm\fr{e^{sb_{j}}H(s)}{2m_{j}(s)K_j^+(s)}+\fr{1}{2\pi i}\int_L \fr{e^{\Gs b_{j}}H(\Gs)d\Gs}{m_{j}(\Gs)K_j^+(\Gs)(\Gs-s)}, \quad 
s\in L,
\label{2.14}
\eeq
and the Cauchy integral is explicitly evaluated by the theory of residues.
Alternatively, this splitting   can be obtained by representing the function $H(s)$ as a sum of $n$ integrals similar to (\ref{2.7})
and then removing the poles. The second approach will be employed in the scalar case
in section \ref{s2.2}.

Now, since the poles of the left- and right-hand sides are known, we apply the Liouville
theorem. Crucial to the success of the method is the fact that the known functions
in the system (\ref{2.13}) are meromorphic functions having a finite number of poles. That is why
the   left- and  right-hand sides are rational functions with prescribed poles and unknown
coefficients. These coefficients can be fixed by the requirement that the final solution $U_j^\pm(s)$
of the vector RHP has to have removable singular points at the poles lying in the half-planes 
$D^\pm$. These conditions form a finite system of linear algebraic equations for the 
unknown coefficients. The case of an order-$2$ RHP when the functions are meromorphic and have an infinite number of  periodic and not periodic  poles will be considered in sections \ref{s3}
and \ref{s4}, respectively. The functions $\Gb_{0j}(p)$ and $\Gb_{0j}(p)$    
introduced in (\ref{2.4}) are fixes by the conditions (\ref{2.3}).

\subsection{Generalization of the Poisson formula for an infinite piece-wise homogeneous rod}\label{s2.2}

To clarify the procedure of finding the functions $\Gb_{0j}(p)$ and $\Gb_{1j}(p)$,
we consider
the heat equation for an infinite rod composed of two semi-infinite rods having different constant diffusivities and conductivities. We assume that the initial temperature, $f_0(x)$, does not vanish
at $\pm\infty$ that is
\beq
f_0(x)=\Gg_-\Gt(-x)+\Gg_+\Gt(x)+f(x), \quad f(x)\in L_1(-\infty,\infty),
\label{2.16}
\eeq
where $\Gg_\pm$ are nonzero constants and $\Gt(x)=1$, $x>0$ and vanishes otherwise.
It is convenient to split the temperature, $u_0(x,t)$ ($|x|<\infty$, $t\ge 0$),  as
\beq
u_0(x,t)=u(x,t)+\Gg_-\Gt(-x)+\Gg_+\Gt(x)
\label{2.17}
\eeq
and determine $u(x,t)$ as the solution of the boundary-value problem
$$
u_t=a^2(x)u_{xx}+g(x,t), \quad |x|<\infty, \quad x\ne 0, \quad t>0,
$$
$$
u|_{x=0^-}-u|_{x=0^+}=\Gg, \quad k_-u_x|_{x=0^-}=k_{j+1}u_x|_{x=0^+}, 
\quad t\ge 0,
$$
\beq
u|_{t=0}=f(x), \quad |x|<\infty,
\label{2.18}
\eeq
where $a(x)=a_-$, $x<0$,   $a(x)=a_+$, $x>0$, and $\Gg=\Gg_+-\Gg_-$. The problem is equivalent
to  the following scalar RHP:
\beq
U^+(s)=-\fr{s^2a_+^2-p}{s^2a_-^2-p}U^-(s)-\fr{H^-(s)+H^+(s)+\Gb_1+s\Gb_0}{s^2a_-^2-p}, \quad s\in L,
\label{2.19}
\eeq
where
$$
\Gb_j=a_-^2\fr{d^j}{dx^j}\hat u(0^-;p)-a_+^2\fr{d^j}{dx^j}\hat u(0^+;p),\quad j=0,1,
$$$$
U^-(s)=\int_0^\infty e^{-sx}\hat u dx, \quad 
U^+(s)=\int_{-\infty}^0 e^{-sx}\hat u dx,
$$
\beq
H^-(s)=\int_0^\infty e^{-sx}[f(x)+\hat g(x;p)] dx, \quad 
H^+(s)=\int_{-\infty}^0 e^{-sx}[f(x)+\hat g(x;p)] dx.
\label{2.20}
\eeq
The functions $U^\pm$ and $H^\pm$ are analytic in the half-planes $D^\pm$. 
The coefficient of the RHP is a rational function, and the functions $U^+(s)$ and $U^-(s)$
are recovered in the standard manner,
$$
U^+(s)=\fr{1}{sa_--\sqrt{p}}
\left[-\fr{a_+h_-}{\sqrt{p}(a_++a_-)}-A_0-\fr{1}{sa_-+\sqrt{p}}\left(
H^+(s)+\fr{a_-h_+(sa_+-\sqrt{p})}{\sqrt{p}(a_++a_-)}\right)\right],
$$
\beq
U^-(s)=\fr{1}{sa_++\sqrt{p}}
\left[\fr{a_-h_+}{\sqrt{p}(a_++a_-)}-A_1-\fr{1}{sa_+-\sqrt{p}}\left(
H^-(s)-\fr{a_+h_-(sa_-+\sqrt{p})}{\sqrt{p}(a_++a_-)}\right)\right],
\label{2.21}
\eeq
where
$$
h_+=H^+\left(-\fr{\sqrt{p}}{a_-}\right), \quad h_-=H^-\left(\fr{\sqrt{p}}{a_+}\right),
$$
\beq
A_0=\fr{\Gb_0\sqrt{p}+a_+\Gb_1}{\sqrt{p}(a_++a_-)},
\quad 
A_1=\fr{\Gb_0\sqrt{p}-a_-\Gb_1}{\sqrt{p}(a_++a_-)},
\label{2.22}
\eeq
Notice that the points $s=\pm\sqrt{p}/a_\pm\in D^\mp$ are removable singularities
of the functions $U^\mp(s)$. The derivation of representations for the functions $\Gb_0(p)$
and $\Gb_1(p)$ requires inversion of the Laplace transforms in (\ref{2.21}). 
This implies
$$
\hat u(x;p)=\left[\fr{2a_-h_+-(a_+-a_-)h_-}{2a_+(a_++a_-)\sqrt{p}}-\fr{A_1}{a_+}\right]
e^{-\sqrt{p}x/a_+}
$$$$
+\fr{1}{2a_+\sqrt{p}}\int_0^\infty[f(\xi)+\hat g(\xi;p)]e^{-\sqrt{p}|x-\xi|/a_+}d\xi,
\quad 0<x<\infty,
$$
$$
\hat u(x;p)=\left[\fr{2a_+h_-+(a_+-a_-)h_+}{2a_-(a_++a_-)\sqrt{p}}+\fr{A_0}{a_-}\right]
e^{\sqrt{p}x/a_-}
$$
\beq
+\fr{1}{2a_-\sqrt{p}}\int_{-\infty}^0[f(\xi)+\hat g(\xi;p)]e^{-\sqrt{p}|x-\xi|/a_-}d\xi,
\quad -\infty<x<0.
\label{2.23}
\eeq
The function $\hat u(x;p)$ and its derivative $\hat u_x(x;p)$ are discontinuous at
the point $x=0$ and due to (\ref{2.18}) have to meet the conditions 
\beq
\hat u|_{x=0^-}-\hat u|_{x=0^+}=\fr{\Gg}{p}, \quad
k_-\hat u_x|_{x=0^-}-k_+\hat u_x|_{x=0^+}=0,
\label{2.24}
\eeq
On satisfying these conditions we eventually determine 
the functions $\Gb_0$ and $\Gb_1$ as
$$
\Gb_0=\fr{1}{a_+k_-+a_-k_+}\left[\fr{(a_-^2-a_+^2)(a_+^2k_-h_++a_-^2k_+h_-}{a_+a_-\sqrt{p}}
+\fr{\Gg(a_+^3k_-+a_-^3k_+)}{p}\right],
$$
\beq
\Gb_1=\fr{a_-^2k_+-a_+^2k_-}{a_+k_-+a_-k_+}
\left[\fr{a_-h_--a_+h_+}{a_+a_-}
+\fr{\Gg}{\sqrt{p}}\right].
\label{2.25}
\eeq
If we substitute these expressions in formulas (\ref{2.23}), the expressions 
for the function $\hat u(x;p)$
are simplified  and become
\beq
\hat u(x;p)=C_\pm(p) e^{\mp\sqrt{p}x/a_\pm}\pm
\fr{1}{2a_\pm\sqrt{p}}\int_0^{\pm\infty}[f(\xi)+\hat g(\xi;p)]e^{-\sqrt{p}|x-\xi|/a_\pm}d\xi,
\quad \pm x>0.
\label{2.26}
\eeq
Here, 
$$
C_+(p)=\fr{\Gl_{1}h_-}{2a_+\sqrt{p}}+\fr{\Gl_{-}h_+}{a_-\sqrt{p}}+\fr{k_-\Gg}{\Gl_0(p-a)}, \quad
C_-(p)=-\fr{\Gl_{1}h_+}{2a_-\sqrt{p}}+\fr{\Gl_{+}h_-}{a_+\sqrt{p}}+\fr{k_+\Gg}{\Gl_0(p+a)},
$$
\beq
\Gl_{1}=\fr{1}{\Gl_0}
\left(\fr{k_+}{a_+}-\fr{k_-}{a_-}\right),\quad 
\Gl_{\pm}=\fr{k_\pm}{a_\pm\Gl_0}, \quad
\Gl_0=\fr{k_+}{a_+}+\fr{k_-}{a_-}.
\label{2.27}
\eeq
To finalize our derivations, we apply the inverse Laplace transform and
take into consideration the formulas
$$
\fr{1}{2\pi i}\int_{c-i\infty}^{c+i\infty}\fr{e^{-\sqrt{p}\Ga+pt}dp}{\sqrt{p}}=\fr{1}{\sqrt{\pi t}}e^{-\Ga^2/(4 t)},
$$
\beq
\fr{1}{2\pi i}\int_{c-i\infty}^{c+i\infty}\fr{e^{-\sqrt{p}\Ga+pt}dp}{p}={\rm Erfc}\left(\fr{\Ga}{2\sqrt{t}}\right),
\quad \R\Ga>0, \quad \R c>0.
\label{2.28}
\eeq
Here, $\rm{Erfc}(\cdot)$ is the  the complementary error function.
This implies the following representations for the function $u(x,t)$ when $x$ is negative:
$$
u(x,t)=\fr{k_+\Gg}{\Gl_0a_+}{\rm Erfc}\left(-\fr{x}{2\sqrt{t}a_-}\right)
+\fr{\Gl_{+}}{a_+\sqrt{\pi t}}
\int_0^{\infty}e^{-(x/a_--\Gx/a_+)^2/(4t)}f(\Gx)d\Gx
$$$$
+\fr{1}{2a_-\sqrt{\pi t}}
\int_{-\infty}^0\left[-\Gl_{1}e^{-(x+\Gx)^2/(4a_-^2t)}+e^{-(x-\Gx)^2/(4a_-^2t)}
\right]f(\Gx)d\Gx
$$
$$
+\fr{\Gl_{+}}{a_+\sqrt{\pi}}\int_0^t\int_0^{\infty}
e^{-(x/a_--\Gx/a_+)^2/[4(t-\tau)]}
\fr{g(\xi,\tau)d\xi d\tau}{\sqrt{t-\tau}}
$$
\beq
+\fr{1}{2a_-\sqrt{\pi}}\int_0^t\int_{-\infty}^0
\left[-\Gl_{1}e^{-(x+\Gx)^2/[4a_-^2(t-\tau)]}+e^{-(x-\Gx)^2/[4a_-^2(t-\tau)]}
\right]
\fr{g(\xi,\tau)d\xi d\tau}{\sqrt{t-\tau}}.
\label{2.29}
\eeq
For $x$ positive we have
$$
u(x,t)=-\fr{k_-\Gg}{\Gl_0a_-}{\rm Erfc}\left(\fr{x}{2\sqrt{t}a_+}\right)
+\fr{\Gl_{-}}{a_-\sqrt{\pi t}}
\int_{-\infty}^0e^{-(x/a_+-\Gx/a_-)^2/(4t)}f(\Gx)d\Gx
$$$$
+\fr{1}{2a_+\sqrt{\pi t}}
\int_0^{\infty}\left[\Gl_{1}e^{-(x+\Gx)^2/(4a_+^2t)}+e^{-(x-\Gx)^2/(4a_+^2t)}
\right]f(\Gx)d\Gx
$$
$$
+\fr{\Gl_{-}}{a_-\sqrt{\pi}}\int_0^t\int_{-\infty}^0
e^{-(x/a_+-\Gx/a_-)^2/[4(t-\tau)]}
\fr{g(\xi,\tau)d\xi d\tau}{\sqrt{t-\tau}}
$$
\beq
+\fr{1}{2a_+\sqrt{\pi}}\int_0^t\int_0^{\infty}
\left[\Gl_{1}e^{-(x+\Gx)^2/[4a_+^2(t-\tau)]}+e^{-(x-\Gx)^2/[4a_+^2(t-\tau)]}
\right]
\fr{g(\xi,\tau)d\xi d\tau}{\sqrt{t-\tau}}.
\label{2.30}
\eeq
The total temperature $u_0(x,t)$ given by formula (\ref{2.17})
is bounded and has different limits as $x\to\pm\infty$.
When $x$ is kept finite and $t\to\infty$, the temperature has a finite limit independent of $x$,
\beq
\lim_{t\to\infty} u_0(x,t)=\fr{\Gg_- a_+k_-+\Gg_+ a_-k_+}{\Gl_0 a_+ a_-}, \quad -X_1<x<X_2.
\label{2.31}
\eeq
Here, $X_1$ and $X_2$ are any finite positive numbers. Formula (\ref{2.31}) is consistent with
the result obtained by Deconinck et al (2014).
If $\Gg_-=\Gg_+$ ($\Gg=0$) and there is no heat source ($g(x,t)\equiv 0$), then
the representations (\ref{2.29}) and (\ref{2.30})
generalize to the discontinuous  case
 the classical Poisson formula obtained for an infinite homogeneous rod.

Notice that it is possible to bypass the  RHP and derive  the representation (\ref{2.26}) for the 
function $\hat u(x;p)$ directly by employing the fundamental functions
\beq
\fr{1}{2a_\pm\sqrt{p}}e^{-\sqrt{p}|x-\Gx|/a_\pm}
\label{2.32}   
\eeq
of the differential operators $a_\pm^2\fr{d^2}{dx^2}-p$. The functions $C_\pm(p)$ are
determined in the same manner as before from the two conditions (\ref{2.24}). Their expressions
coincide with those given by (\ref{2.27}). It is evident that the same approach works for any
number of discontinuities including the case of a finite discontinuous rod with any
physical boundary conditions imposed at the ends. In this case 
the fundamental functions (\ref{2.32}) need to be replaced by the corresponding 
Green functions of the one-dimensional boundary value problems. These Green functions
are derived in an elementary fashion.
If the number of discontinuities is $n\ge 3$, then instead of two functions $C_+(p)$ and $C_-(p)$
we have $n-1$ pairs of unknown functions. They are determined by a system of $2n-2$
linear algebraic equations following from the $2n-2$ conditions at the discontinuity points.

\subsection{Abrahams-Wickham system of integral equations}

Abrahams and Wickham (1990) analyzed the system 
\beq
u(x)=\Gl\int_0^\infty k(x-t)u(t)dt+f(x), \quad 0<x<\infty,
\label{W.1}
\eeq
where the matrix-kernel is given by  
\beq
k(x)=\left(\begin{array}{cc}
e^{-|x|} & e^{-|x-a|}\\
e^{-|x+a|} & e^{-|x|}\\
\end{array}\right),
\label{W.2}
\eeq
$\Gl$ and $a$ are parameters, $u(x)=(u_1(x),u_2(x))^\top$, and $f(x)$
is a forcing vector-function prescribed accordingly. To factorize the matrix coefficient of the RHP associated with the system (\ref{W.1}),
they expressed the matrix-factors through the solution of a certain auxiliary
system of integral equations. In the case (\ref{W.2}) that system 
admits an exact solution. In what follows we derive a closed-form 
solution by a simple method that bypasses not only the auxiliary system
of integral equations, but also the matrix Wiener-Hopf factorization.
First we apply the Fourier integral transform to the system (\ref{W.1})
and have the following RHP on the real axis
\beq
G(\Ga)U^+(\Ga)=U^-(\Ga)+F^+(\Ga), \quad -\infty<\Ga<+\infty,
\label{W.3}
\eeq
where
$$
G(\Ga)=\fr{1}{\Ga^2+1}\left(\begin{array}{cc}
\Ga^2+1-2\Gl  & -2\Gl e^{i\Ga a}\\
-2\Gl e^{-i\Ga a} & \Ga^2+1-2\Gl  \\
\end{array}\right),
\quad F^+(\Ga)=\int_0^\infty f(x) e^{i\Ga x} dx,
$$
\beq
U^+(\Ga)=\int_0^\infty u(x) e^{i\Ga x} dx,\quad
U^-(\Ga)=\int_{-\infty}^0 u_-(x) e^{i\Ga x} dx,
\label{W.4}
\eeq
the vector-functions $U^\pm(\Ga)=(U_1^\pm(\Ga),U_2^\pm(\Ga))^\top$ are analytic in the half-planes $C^\pm=\pm\I\Ga>0$, the vector-function 
 $F^+(\Ga)=(F_1^+(\Ga),F_2^+(\Ga))^\top$ is analytic in the upper half-plane $C^+$, and the vector-function $u_-(x)$, $x<0$, is given by $u_-(x)=\Gl\int_0^\infty k(x-t)u(t)dt$.

Next, instead of factorizing the matrix $G(\Ga)$ we express the function $U_1^+(\Ga)$ from the first equation in (\ref{W.3}) and substitute it into the second equation. After obvious simplifications this brings us
to the new system of functional equations
$$
\fr{\Ga^2+1-2\Gl}{\Ga^2+1}U_1^+(\Ga)-\fr{2\Gl e^{i\Ga a}}{\Ga^2+1}U_2^+(\Ga)=
F_1^+(\Ga)+U_1^-(\Ga),
$$
\beq
\fr{\Ga^2+1-2\Gl}{\Ga^2+1-4\Gl}[U_2^-(\Ga)+F_2^+(\Ga)]
+\fr{2\Gl e^{-i\Ga a}}{\Ga^2+1-4\Gl}[U_1^-(\Ga)+F_1^+(\Ga)]=
U_2^+(\Ga).
\label{W.5}
\eeq
Notice that the products $e^{i\Ga a}U_2^+(\Ga)$ and 
$e^{-i\Ga a}U_1^-(\Ga)$ are analytic in the domains $C^+$
and $C^-$, respectively, and vanish exponentially when $\Ga\to\infty$ and $\Ga\in C^\pm$.
For simplicity, we analyze further the normal case that is
we assume that $\Gl\notin[1/4,+\infty)$. 
Choose $\arg(1-4\Gl)\in(-\pi,\pi)$, $\arg(1-2\Gl)\in(-\pi,\pi)$,
and denote $\Gl_0=\sqrt{1-2\Gl}$,  $\Gl_1=\sqrt{1-4\Gl}$,
$\arg\Gl_j\in (-\pi/2,\pi/2)$, $j=0,1$. Factorize now the rational functions
\beq
\fr{\Ga^2+1-2\Gl}{\Ga^2+1}=\fr{K_0^+(\Ga)}{K_0^-(\Ga)}, \quad 
\fr{\Ga^2+1-2\Gl}{\Ga^2+1-4\Gl}=\fr{K_1^+(\Ga)}{K_1^-(\Ga)},
\label{W.6}
\eeq
where
\beq
K_0^\pm(\Ga)=\left(\fr{\Ga\pm i\Gl_0}{\Ga\pm i}\right)^{\pm 1},\quad
K_1^\pm(\Ga)=\left(\fr{\Ga\pm i\Gl_0}{\Ga\pm i\Gl_1}\right)^{\pm 1}.
\label{W.7}
\eeq
In the general case of the forcing vector-function $f(x)$ we need to introduce new functions, $\GY_1(\Ga)$ and $\GY_2(\Ga)$, the Cauchy integrals
with the density chosen according to the following  relations
for their limit values on the real axis
$$
\GY_1^+(\Ga)-\GY_1^-(\Ga)=F_1^+(\Ga)K_0^-(\Ga),\quad -\infty<\Ga<+\infty,
$$
\beq
\GY_2^+(\Ga)-\GY_2^-(\Ga)=\fr{F_2^+(\Ga)}{K_1^-(\Ga)}
+\fr{2\Gl e^{-i\Ga a} 
F_1^+(\Ga)}{(\Ga+i\Gl_0)(\Ga-i\Gl_1)},\quad -\infty<\Ga<+\infty.
\label{W.8}
\eeq
We seek the solution $U_j^\pm(\Ga)$ in the class of functions vanishing at infinity. Therefore the subsequent application of the generalized Liouville theorem  enables us to determine the solution
of the RHP in the form
$$
U_1^-(\Ga)=\fr{1}{K_0^-(\Ga)}\left[\Psi_1^-(\Ga)+\fr{C_1}{\Ga-i\Gl_0}\right],
$$$$
U_2^+(\Ga)=K_1^+(\Ga)\left[\Psi_2^+(\Ga)+\fr{C_2}{\Ga+i\Gl_0}\right],
$$$$
U_2^-(\Ga)=K_1^-(\Ga)\left[\Psi_2^-(\Ga)+\fr{C_2}{\Ga+i\Gl_0}-\fr{2\Gl e^{-ia\Ga}}{(\Ga+i\Gl_0)(\Ga-i\Gl_1)K_0^-(\Ga)}\left(\Psi_1^-(\Ga)+\fr{C_1}{\Ga-i\Gl_0}\right)\right],
$$
\beq
U_1^+(\Ga)=\fr{1}{K_0^+(\Ga)}\left[\Psi_1^+(\Ga)+\fr{C_1}{\Ga-i\Gl_0}+\fr{2\Gl e^{ia\Ga}K_1^+(\Ga)}{(\Ga+i)(\Ga-i\Gl_0)}\left(\Psi_2^+(\Ga)+\fr{C_2}{\Ga+i\Gl_0}\right)\right].
\label{W.9}
\eeq
Here, $C_1$ and $C_2$ are arbitrary constants. It is easy to observe that the functions
$U_1^+(\Ga)$ and $U_2^-(\Ga)$ have inadmissible simple poles 
at the points $\Ga=i\Gl_0\in C^+$ and  $\Ga=-i\Gl_0\in C^-$, respectively. 
They can be removed if the residues of the functions $U_1^+(\Ga)$ and $U_2^-(\Ga)$ 
at these points vanish. Simple calculations 
give the following expressions for the constants $C_1$ and $C_2$:
\beq
C_1=\fr{d_1+bd_2}{b^2+1}, \quad C_2=\fr{d_2-bd_1}{b^2+1}, 
\label{W.10}
\eeq
where
$$
b=\fr{2\Gl e^{-a\Gl_0}}{(\Gl_0+1)(\Gl_0+\Gl_1)},
$$
\beq
d_1=2i b\Gl_0\Psi_2^+(i\Gl_0),\quad d_2=2i b\Gl_0\Psi_1^-(-i\Gl_0).
\label{W.11}
\eeq
The inverse Fourier transformation of formulas (\ref{W.9}) for $U_1^+(\Ga)$
and  $U_2^+(\Ga)$ yields the solution of the original system of integral equations
(\ref{W.1}).

\setcounter{equation}{0}

\section{Laplace equation in a wedge with mixed boundary conditions: 
the case of meromorphic
functions with periodic poles and zeros}\label{s3}

As an illustration of the method in the case when the 
entries of the matrix coefficient of the RHP
are almost periodic functions and meromorphic functions with periodic poles and zeros,
we consider the following mixed boundary value problem for the Laplace operator:
$$
\GD u(r,\Gt)=0, \quad 0<r<\infty, \quad 0<\Gt<\Ga,
$$$$
u|_{\Gt=0}=f_{1-}(r), \quad 0<r<a_1; \quad
\left. -\fr{\Md u}{r\Md\Gt}\right|_{\Gt=0}=f_{1+}(r), \quad r>a_1,
$$
\beq
u|_{\Gt=\Ga}=f_{2-}(r), \quad 0<r<a_2; \quad \left. \fr{\Md u}{r\Md\Gt}\right|_{\Gt=\Ga}=f_{2+}(r), \quad r>a_2.
\label{3.1}
\eeq
The function $u$ is sought in the class of functions bounded at $r=0$ and vanishing
at $r=\infty$ as $|u(r,\Gt)|\le Ar^{-\Gb}$,  $0\le\Gt\le\Ga$, $A=\const$,   $0<\Gb<\Gve$, 
and $\Gve$
is a small positive number.
 This problem can be interpreted as a model of stationary heat conduction
if $u^{(t)}(r,\Gt)=u(r,\Gt)+T_\infty$ is the temperature and $T_\infty$ is the 
temperature
at infinity. $T_\infty$ is constant for all $\Gt\in[0,\Ga]$ and has to be determined {\it a posteriori}.
The temperature $u^{(t)}=T_1$ is prescribed in the segment $0<r<a_1$, $\Gt=0$ as
$T_1(r)=f_{1-}(r)+T_\infty$ and in the segment  $0<r<a_2$, $\Gt=\Ga$ as
$u^{(t)}=T_2(r)=f_{2-}(r)+T_\infty$. In the rest of the boundary, the heat flux $q$ is
given: $q=k_-f_{1+}(r)$, $r>a_1$, $\Gt=0$, and  $q=k_+f_{2+}(r)$, $r>a_2$, $\Gt=\Ga$.
($k_-$ and $k_+$ are the thermal conductivities of the lower and upper boundaries, respectively).
It is also assumed that the functions $f_{j+}(r)$ decay at infinity as
$f_{j+}(r)=O(r^{-1-\Gb})$, $r\to\infty$, $j=1,2$.

\subsection{Vector RHP}

Before proceeding with the solution, we extend all  the boundary conditions for the whole 
semi-axis as
$$
u|_{\Gt=0}=f_{1-}(r)+\Gvf_{1+}(r), \quad
\left. -\fr{\Md u}{r\Md\Gt}\right|_{\Gt=0}=\Gvf_{1-}(r)+f_{1+}(r), \quad 0<r<\infty,
$$
\beq
u|_{\Gt=\Ga}=f_{2-}(r)+\Gvf_{2+}(r) , \quad \left. \fr{\Md u}{r\Md\Gt}\right|_{\Gt=\Ga}=\Gvf_{2-}(r)+f_{2+}(r), \quad 0<r<\infty,
\label{3.2}
\eeq
where
$$
\supp f_{j-}(r)\subset[0,a_j], \quad \supp f_{j+}(r)\subset[a_j,\infty),
$$
\beq
\supp \Gvf_{j-}(r)\subset[0,a_j], \quad \supp \Gvf_{j+}(r)\subset[a_j,\infty),
\label{3.3}
\eeq
and the functions $\Gvf_{j\pm}(r)$ need to be recovered.
We shall now apply the Mellin transformation  to the Laplace equation
and the extended boundary conditions (\ref{3.2}). Let
$$
F_j^-(s)=\int_0^1 f_j(a_j\Gr)\Gr^{s-1}d\Gr, \quad F_j^+(s)=a_j\int_1^\infty 
f_j(a_j\Gr)\Gr^{s}d\Gr, 
$$
\beq
\GF_j^-(s)=a_j\int_0^1 \Gvf_{j-}(a_j\Gr)\Gr^{s}d\Gr, \quad \GF_j^+(s)=\int_1^\infty 
\Gvf_{j+}(a_j\Gr)\Gr^{s-1}d\Gr, 
\label{3.4}
\eeq
and 
\beq
\hat u_s(\Gt)=\int_0^\infty u(r,\Gt)r^{s-1}dr, \quad s\in L: \R s=\Gs\in(0,\Gb).
\label{3.5}
\eeq
The function $\hat u_s(\Gt)$ solves the differential equation $\hat u_s''(\Gt)+s^2 \hat u_s(\Gt)=0$
and has the form
\beq
\hat u_s(\Gt)=A_1(s)\cos s\Gt+A_2(s)\sin s\Gt,
\label{3.6}
\eeq
where $A_1$ and $A_2$ are fixed from the first two boundary conditions in (\ref{3.2}) as
\beq
A_1=a_1^s[\GF_1^+(s)+F_1^-(s)], \quad A_2=-\fr{a_1^s}{s}[\GF_1^-(s)+F_1^+(s)].
\label{3.7}
\eeq
The third and fourth boundary conditions in  (\ref{3.2})  constitute the vector RHP
\beq
\GF^+(s)=-\fr{1}{s}G(s)[\GF^-(s)+F^+(s)]-F^-(s), \quad s\in L,
\label{3.8}
\eeq
where
\beq
G(s)=\left(\begin{array}{cc}
\cot \Ga s & \Gl^s\csc \Ga s\\
 \Gl^{-s}\csc \Ga s & \cot \Ga s\\
 \end{array}
 \right), \quad \Gl=\fr{a_2}{a_1}.
 \label{3.9}
 \eeq
 Without loss of generality we assume that $\Gl>1$. The vectors $\GF^\pm(s)=(\GF_1^\pm(s),
 \GF_2^\pm(s))^\top$ are analytic in the half-planes $D^\pm: \pm \R s<\pm\Gs$, 
$ \Gs\in(0,\Gb)$, and $0\in D^+$.

\subsection{Factorization of the matrix $G(s)$}

We aim to find two matrices, $X^+(s)$ and $X^-(s)$, analytic in the domains $D^+$ and $D^-(s)$,
respectively, having a finite order at infinity and solving the following matrix equation
\beq
X^+(s)=G(s)X^-(s), \quad s\in L.
\label{3.10}
\eeq
Denote
\beq
X^\pm(s)=\left(\begin{array}{cc}
\Gc_{11}^\pm(s)  & \Gc_{12}^\pm(s)\\
 \Gc_{21}^\pm(s) & \Gc^\pm_{22}(s)\\
 \end{array}
 \right).
 \label{3.10}
 \eeq
Regrouping terms in the same fashion as in section 2.1 we obtain
$$
\Gc_{1j}^+(s)=-\tan\Ga s\Gc_{1j}^-(s)+\fr{\Gl^s\Gc_{2j}^+(s)}{\cos\Ga s},
$$
\beq
\Gc_{2j}^-(s)=\tan\Ga s\Gc_{2j}^+(s)-\fr{\Gl^{-s}\Gc_{1j}^-(s)}{\cos\Ga s}.
\label{3.11}
\eeq
It is seen that the functions $\Gc^+_{1j}(s)$ and  $\Gc^-_{2j}(s)$ have 
simple poles  at the points  $s=-\fr{\pi}{\Ga}(n+\fr12)\in D^+$ and
 $s=\fr{\pi}{\Ga}(n+\fr12)\in D^-$ ($n=0,1,\ldots$), respectively, and they have to be removed. 
 We seek the functions
 $\Gc_{1j}^-(s)$ and $\Gc_{2j}^+(s)$ in the form of the hypergeometric series
 $$
 \Gc^-_{1j}(s)=\sum_{k=0}^\infty\fr{\GG(\Ga s/\pi+k+\nu_1)\GG(k+\nu_2)}
{\GG(\Ga s/\pi+k+\mu_1+1/2)\GG(k+\mu_2)}(-1)^k\Gl^{-k\pi/\Ga+\Gs_1},
$$
\beq
\Gc^+_{2j}(s)=\Gk\sum_{k=0}^\infty\fr{\GG(-\Ga s/\pi+k+\Ga_1)\GG(k+\Ga_2)}
{\GG(-\Ga s/\pi+k+\Gb_1+1/2)\GG(k+\Gb_2)}(-1)^k\Gl^{-k\pi/\Ga+\Gs_2}.
\label{3.12}
\eeq
The parameters  $\nu_j$, $\mu_j$, $\Ga_j$, $\Gb_j$, $\Gs_j$ ($j=1,2$), and $\Gk$ are to be determined 
from the following conditions:

(i) the functions $\Gc^+_{1j}(s)$ have removable singularities at the zeros
of $\cos\Ga s$ lying in the half-plane $D^+$,
\beq
-\sin\Ga s \Gc_{1j}^-(s)+\Gl^s\Gc_{2j}^+(s)=0, \quad s=-\fr{\pi}{\Ga}\left( n+\fr12\right),\quad
n=0,1,\ldots,
\label{3.13}
\eeq

(ii)  the functions $\Gc^-_{2j}(s)$ have removable singularities at the points
 $s=\fr{\pi}{\Ga}(n+\fr12)\in D^-$ 
\beq
\sin\Ga s \Gc_{2j}^+(s)-\Gl^{-s}\Gc_{1j}^-(s)=0, \quad s=\fr{\pi}{\Ga}\left( n+\fr12\right),\quad
n=0,1,\ldots,
\label{3.14}
\eeq

(iii) the functions $\Gc^-_{1j}(s)$ may have
simple poles at the zeros of  $\tan\Ga s$ lying in $D^+$,
$s=-\pi n/\Ga$, $n=0,1,\ldots$; it follows from (\ref{3.11}) that then the functions 
$\Gc_{1j}^+(s)$ have removable singularities at these points,

(iv) the functions $\Gc^+_{2j}(s)$ may have
simple poles at the points $s=\pi n/\Ga\in D^-$,
$n=1,2,\ldots$. Then the functions 
$\Gc_{2j}^-(s)$ have removable singularities at these points.

The conditions (i) and (ii) when satisfied give 
$$
\Gb_2=1, \quad \Gm_2=1, \quad \Gb_1=1-\mu_1, \quad \Ga_2=1/2+\nu_1-\mu_1,
$$
\beq
\nu_2=\Ga_1+\mu_1-1/2,\quad \Gk=(-1)^{\mu_1}, \quad \Gs_2=\Gs_1+\fr{\pi(2\mu_1-1)}{2\Ga}.
\label{3.15}
\eeq
The other two conditions, (iii) and (iv), determine $\nu_1$ and $\Ga_1$:
$\nu_1=0$ and $\Ga_1=1$. Without loss of generality, $\Gs_1=0$. This brings us to the following one-parametric family of solutions:
$$
 \Gc^-_{1j}(s)=\sum_{k=0}^\infty\fr{\GG(\Ga s/\pi+k)\GG(k+\mu_1+1/2)(-1)^k
 \Gl^{-k\pi/\Ga}}
{\GG(\Ga s/\pi+k+\mu_1+1/2)k!},
$$
\beq
\Gc^+_{2j}(s)=\sum_{k=0}^\infty\fr{\GG(-\Ga s/\pi+k+1)\GG(k-\mu_1+1/2)(-1)^{k-\mu_1}
 \Gl^{(\mu_1-k-1/2)\pi/\Ga}}
{\GG(-\Ga s/\pi+k-\mu_1+3/2)k!}.
\label{3.16}
\eeq
On choosing $\mu_1=0$ for $j=1$ we obtain the functions $\Gc_{11}^-(s)$
 and $\Gc_{21}^+(s)$ vanishing at infinity as $\Gc_{11}^-(s)=O(s^{-1/2})$,
 $s\in D^-$, and   $\Gc_{21}^+(s)=O(s^{-1/2})$,
 $s\in D^+$, and admitting  the following representations through
 the beta and Gauss functions:
 $$
\Gc_{11}^-(s)=B\left(\fr{\Ga s}{\pi},\fr12\right)F\left(
\fr{\Ga s}{\pi},\fr12;\fr{\Ga s}{\pi}+\fr12;-\Gl^{-\pi/\Ga}
\right),
$$
\beq
\Gc_{21}^+(s)=\Gl^{-\pi/(2\Ga)}B\left(-\fr{\Ga s}{\pi}+1,\fr12\right)F\left(
-\fr{\Ga s}{\pi}+1,\fr12;-\fr{\Ga s}{\pi}+\fr32;-\Gl^{-\pi/\Ga}
\right).
\label{3.17}
\eeq
The function $\Gc_{11}^-(s)$ is analytic in the half-plane $D^-$ and meromorphic
in $D^+$, while the function $\Gc_{21}^+(s)$ is analytic in $D^+$ and 
meromorphic in $D^-$. In the case $j=2$, we put $\mu_1=1$ and derive
the other two functions
$$
\Gc_{12}^-(s)=B\left(\fr{\Ga s}{\pi},\fr32\right)F\left(
\fr{\Ga s}{\pi},\fr32;\fr{\Ga s}{\pi}+\fr32;-\Gl^{-\pi/\Ga}
\right),
$$
\beq
\Gc_{22}^+(s)=-\Gl^{\pi/(2\Ga)}B\left(-\fr{\Ga s}{\pi}+1,-\fr12\right)F\left(
-\fr{\Ga s}{\pi}+1,-\fr12;-\fr{\Ga s}{\pi}+\fr12;-\Gl^{-\pi/\Ga}
\right).
\label{3.18}
\eeq 
Notice that the function $\Gc_{12}^-(s)$ vanishes at infinity, while the function 
$\Gc_{22}^+(s)$ grows as $s\to\infty$:
$\Gc_{12}^-(s)=O(s^{-3/2})$,  $\Gc_{22}^+(s)=O(s^{1/2})$.

The functions $\Gc_{1j}^+(s)$ and $\Gc_{2j}^-(s)$
are expressed through the functions (\ref{3.17}) and  (\ref{3.18}) by formulas (\ref{3.11}).
They are analytic in $D^+$ and $D^-$, and their order at infinity is determined by the order of $\Gc_{1j}^-(s)$ and $\Gc_{2j}^+(s)$, respectively. This completes the exact factorization
of the matrix $G(s)$. The factorization we found is not unique:
by choosing the parameter $\mu_1$ as an integer different from 0 and 1 we can
construct another set of the 
functions  $\Gc_{12}^-(s)$ and $\Gc_{22}^+(s)$ and that is why another factorization
different from the one found. The Wiener-Hopf matrix factors $X^+(s)$ and $X^-$ given by 
(\ref{3.17})
and (\ref{3.18})
have the following asymptotics at infinity:
\beq
X^\pm(s)=\left(\begin{array}{cc}
O(s^{-1/2})  & O(s^{-3/2})\\
O(s^{-1/2}) & O(s^{1/2})\\
 \end{array}
 \right), \quad s\to\infty.
 \label{3.19}
 \eeq

\subsection{Exact solution of the heat conduction problem in a wedge}

On having factorized the matrix $G(s)$ we replace  $G(s)$ by
the product $X^+(s)[X^-(s)]^{-1}$ in the boundary condition (\ref{3.8})
of the vector RHP
\beq
[X^+(s)]^{-1}[\GF^+(s)+F^-(s)]=-\fr{1}{s}[X^-(s)]^{-1}[\GF^-(s)+F^+(s)], \quad s\in L.
\label{3.20}
\eeq
Represent next the functions $f_{1-}(r)$ and $f_{2-}(r)$
as
\beq
f_{j-}(r)=\tilde T_j+T_j^*(r), \quad \tilde T_j=-T_\infty+T_j^\circ, \quad j=1,2,
\label{3.21}
\eeq
where $T_j^\circ+T_j^*(r)=T_j(r)$ is the prescribed temperature in the lower
boundary $0<r<a_1$ and the upper boundary $0<r<a_2$ of the wedge,
$T_j^\circ=\const$, and $T_j^*(r)=O(r^{\Gg_j})$, $r\to 0$, $\Gg_j>0$.
We also introduce the Mellin transforms
\beq
\hat T_j^-(s)=\int_0^1 T_j^*(a_j\Gr)\Gr^{s-1}d\Gr, \quad j=1,2,
\label{3.22}
\eeq
and the Cauchy integrals
$$
\Psi(s)=\fr{1}{2\pi i}\int_L\fr{[X^+(\Gs)]^{-1}\hat T^-(\Gs)d\Gs}{\Gs-s},
$$
\beq
\GO(s)=\fr{1}{2\pi i}\int_L\fr{[X^-(\Gs)]^{-1} F^+(\Gs)d\Gs}{\Gs(\Gs-s)}.
\label{3.23}
\eeq
Employ now the continuity principle and the Liouville theorem and derive the following
representations of the solution:
$$
\GF^+(s)=\left(-I+X^+(s)[X^+(0)]^{-1}\right)\fr{\tilde T}{s}-X^+(s)[\Psi^+(s)+\GO^+(s)],
$$
\beq
\GF^-(s)=-X^-(s)[X^+(0)]^{-1}\tilde T+sX^-(s)[\Psi^-(s)+\GO^-(s)],
\label{3.24}
\eeq
where $\tilde T=(\tilde T_1,\tilde T_2)^\top$. Further, since the functions $\Gc^\pm_{22}(s)$
are growing at infinity, and the 
matrix factors have the  asymptotics (\ref{3.19}), the second components
of the vectors $\GF^-(s)$ and $\GF^+(s)$, in general,  have the 
asymptotics $\GF_2^-(s)=O(s^{1/2})$,  $\GF_2^+(s)=O(s^{-1/2})$.
Due to the Tauberian theorems for the Mellin transforms this causes
an infinite temperature as $r\to a_j^+$ at the lower ($j=1$) and the upper ($j=2$)
boundaries. Also, as $r\to a_j^-$, the heat flux has a nonintegrable singularity
of order $3/2$.
To derive the condition which guaranties the boundedness of the temperature and integrability
of the heat flux, we denote the limits
$$
\Psi^\circ=(\Psi_1^\circ,\Psi_2^\circ)^\top=\lim_{s\to\infty, s\in D^\pm}s\Psi^\pm(s),
$$
\beq
\GO^\circ=(\GO_1^\circ,\GO_2^\circ)^\top=\lim_{s\to\infty, s\in D^\pm}s\GO^\pm(s),
\label{3.25}
\eeq
and compute $[X^+(0)]^{-1}\tilde T$. It is directly verified that 
\beq
\Gc_{11}^+(0)=-\pi+\Gc_{21}^+(0), \quad 
\Gc_{12}^+(0)=-\pi+\Gc_{22}^+(0), 
\label{3.26}
\eeq
and
\beq
\Gc_{21}^+(0)=2\tan^{-1}\Gl^{-\pi/(2\Ga)}, \quad 
\Gc_{22}^+(0)=2\Gl^{\pi/(2\Ga)}+2\tan^{-1}\Gl^{-\pi/(2\Ga)}.
\label{3.27}
\eeq
Continuing, the vectors $\GF^\pm(s)$ have the asymptotics at infinity we need
if and only if 
\beq
-\tilde T_1+\tilde T_2\left(1-\fr{\pi}{\Gc_{21}^+(0)}\right)=\fr{\GD^+(0)}{\Gc_{21}^+(0)}
(\Psi_2^\circ+\GO_2^\circ),
\label{3.28}
\eeq
where $\GD^+(0)=\det X^+(0)=-2\pi\Gl^{\pi/(2\Ga)}$. If this condition is fulfilled, then 
as $s\to\infty$,
$\GF^-(s)=O(s^{-1/2})$, $s\in D^-$, and  $\GF^+(s)=O(s^{-1})$, $s\in D^+$.
The condition can be easily satisfied by the corresponding choice of the  parameter  $T_\infty$
undetermined at this stage. Since $\tilde T_j=-T_\infty+T_j^\circ$, we transform equation 
(\ref{3.28})  and find the temperature at infinity
\beq
T_\infty=\fr{2}{\pi}\tan^{-1}\Gl^{-\pi/(2\Ga)}(T_1^\circ-T_2^\circ)+T_2^\circ-2\Gl^{\pi/(2\Ga)}
(\Psi_2^\circ+\GO_2^\circ).
\label{3.29}
\eeq
In particular, if $T_j^*(r)=0$ and $f_{j+}(r)=0$, then $\Psi_2^\circ=0$ and $\GO^\circ_2=0$.
On assuming further that $T_1^\circ=0$ (the case considered by Abrahams and Wickham (1990)),
we obtain the simple explicit formula 
\beq
T_\infty=\left[-\fr{2}{\pi}\tan^{-1}\Gl^{-\pi/(2\Ga)}+1\right]T_2^\circ.
\label{3.30}
\eeq
Finally, if $a_1=a_2$, then $\Gl=1$ and therefore $T_\infty=\fr12 T_2^\circ$.
The same result for the case $\Gl=1$ was derived by Abrahams and Wickham (1990).

We wish also to recover the function $u(r,\Gt)$ and study its behavior as
$r\to\infty$ and $r\to 0$.
On applying the inverse Mellin transform to (\ref{3.6}) and using (\ref{3.7}) and (\ref{3.8})
it is possible to have
$$
u(r,\Gt)=-\fr{1}{2\pi i}\int_L\left\{[\GF_1^-(s)+F_1^+(s)]\cos(\Ga-\Gt)s\left(\fr{r}{a_1}\right)^{-s}
\right.
$$
\beq
\left.+[\GF_2^-(s)+F_2^+(s)]\cos\Gt s\left(\fr{r}{a_2}\right)^{-s}\right\}\fr{ds}{s\sin\Ga s}.
\label{3.31}
\eeq
Denote by $\Gk_j$ those singular points of the functions $F_j^+(s)$ 
in the domain $D^-$ which have the smallest  real part among
the singular points of $F_j^+(s)$ in $D^-$.
Then we can derive
\beq
u(r,\Gt)=O(r^{-\Gb}), \quad r\to\infty, \quad \Gb=\min\{\pi/\Ga,\Gk_1,\Gk_2\}>0.
\label{3.32}
\eeq
In particular, when $f_{1+}(r)=f_{2+}(r)=0$ and $r\to\infty$,
\beq
u(r,\Gt)=\fr{1}{\pi}\cos\fr{\pi\Gt}{\Ga}
\left[\GF_1^-\left(\fr{\pi}{\Ga}\right)\left(\fr{r}{a_1}\right)^{-\pi/\Ga}-\GF_2^-\left(\fr{\pi}{\Ga}\right)\left(\fr{r}{a_2}\right)^{-\pi/\Ga}
\right]
+O(r^{-2\pi/\Ga}), \quad
\label{3.33}
\eeq
and as $\Gt=\Ga/2$, $u\sim cr^{-2\pi/\Ga}$, $r\to\infty$, $c$ is a nonzero constant.

If we want to determine the behavior of the function $u$ as $r\to 0$, we need regroup
the terms in the integrand (\ref{3.31}) in order to get rid of the functions $\GF^-_1(s)$
and $\GF_2^-(s)$. We have
$$
u(r,\Gt)=\fr{1}{2\pi i}\int_L\left\{[\GF_1^+(s)+F_1^-(s)]\sin(\Ga-\Gt)s\left(\fr{r}{a_1}\right)^{-s}
\right.
$$
\beq
\left.+[\GF_2^+(s)+F_2^-(s)]\sin\Gt s\left(\fr{r}{a_2}\right)^{-s}\right\}\fr{ds}{\sin\Ga s}.
\label{3.34}
\eeq
Accepting the representations (\ref{3.21}) by the Cauchy theorem we derive
\beq
u(r,\Gt)\sim\left(1-\fr{\Gt}{\Ga}\right)T_1^\circ+\fr{\Gt}{\Ga}T_2^\circ-T_\infty, \quad r\to 0.
\label{3.35}
\eeq
Finally, we notice that  the solution we obtained have the following asymptotics 
at the points where the type of the boundary conditions is changed:
\beq
u\sim c_j, \quad r\to a_j^+, \quad
\fr{\Md u}{\Md\Gt}=O(r^{-1/2}),  \quad r\to a_j^-, \quad \Gt=\left\{\begin{array}{cc}
\Gt=0, & j=1,\\
\Gt=\Ga, & j=2.\\
\end{array}
\right.
\label{3.36}
\eeq

\setcounter{equation}{0}

\section{Helmholtz equation in a strip:  nonperiodic poles and zeros}\label{s4}

With the example of the Helmholtz equation in a strip with a cut inside
we shall show how the method of the RHP can be generalized
when the zeros and poles of the meromorphic entries of the RHP matrix coefficient
are not periodic.
Consider the Neumann boundary value problem for the Helmholtz equation
in the doubly connected domain $\GP\setminus S$ with $\GP$ being a strip, $\GP=\{|x_1|<\infty, 
-b_-^\circ<x_2<b_+^\circ\}$, and 
$S$ being a cut,
 $S=\{0<x_1<a, x_2=0^\pm\}$,
$$
(\GD+k_0^2)u^\circ=0, \quad (x_1,x_2)\in \GP\setminus S,
$$
$$
\fr{\Md u^\circ}{\Md x_2}=0, \quad  |x_1|<\infty, \quad x_2=-b_-^\circ, b_+^\circ,
$$
\beq
\fr{\Md u^\circ}{\Md x_2}=f^\circ(x_1), \quad 0<x_1<a, \quad x_2=0^\pm.
\label{4.1}
\eeq
This problem can be interpreted as an antiplane  problem for a strip $\Pi$
with a crack $S$ when the strip boundary is free of traction and the crack faces are subjected 
to the oscillating load $\tau_{23}=e^{i\Go t}Gf^\circ$, $t$ is time, $\Go$ is the frequency, and $G$ is the shear modulus.
This problem also describes acoustic wave propagation in a guideline with an acoustically
hard screen $S$ inside.
 
We assume that $\I k_0>0$.
It will be convenient to work in the dimensionless coordinates 
$x=x_1/a$, $y=x_2/a$. Denote $k=ak_0$, 
$b_\pm=b_\pm^\circ/a$, 
$f(x)=af^\circ(ax)$, 
$u(x,y)=u^\circ(ax,ay)$.
This transforms the problem to
$$
(\GD+k^2)u=0, \quad (x,y)\in \{|x|<\infty, -b_-<y<b_+\}\setminus \{0<x<1, y=0^\pm\},
$$
$$
\fr{\Md u}{\Md y}=0, \quad  |x|<\infty, \quad y=\pm b_\pm,
$$
\beq
\fr{\Md u}{\Md y}=f(x), \quad 0<x<1, \quad y=0^\pm, \quad
u(x,0^+)-u(x,0^-)=\Gvf_1(x), \quad |x|<\infty,
\label{4.2}
\eeq
where $\supp\Gvf_1\subset[0,1]$ and $\Gvf_1(x)$ is an unknown function in the segment $[0,1]$.

On applying the Fourier transform with respect to $x$ 
\beq
\hat u_\Ga(y)=\int_{-\infty}^\infty u(x,y)e^{i\Ga x}dx
\label{4.3}
\eeq
to the Helmholtz equation we easily obtain 
\beq
\hat u_\Ga(y)=\left\{
\begin{array}{cc}
C_1^+\cosh\Gg(y-b_+)+C_2^+\sinh\Gg(y-b_+), & 0<y<b_+,\\
C_1^-\cosh\Gg(y+b_-)+C_2^-\sinh\Gg(y+b_-), & -b_-<y<0,\\
\end{array}
\right.
\label{4.5}
\eeq
where $\Gg=\sqrt{\Ga^2-k^2}$ is the single branch of the algebraic function
$\Gg^2=\Ga^2-k^2$ fixed by the condition $\R\Gg\ge 0$ or, equivalently,
$\Gg(0)=-i$ in the $\Ga$-plane cut along the straight line joining
the branch points $k$ and $-k$ and passing through the infinite point.
The boundary conditions on the sides $y=\pm b_\pm$ and the discontinuity of 
$u$ and continuity of its normal derivative in the line $y=0$ yield
\beq
C_2^+=C_2^-=0, \quad C_1^+=\fr{\sinh\Gg b_-\GF_1^+(\Ga)}{\sinh\Gg(b_++b_-)},
\quad
C_1^-=-\fr{\sinh\Gg b_+\GF_1^+(\Ga)}{\sinh\Gg(b_++b_-)},
\label{4.6}
\eeq
where
\beq
\GF_1^+(\Ga)=\int_0^1\Gvf_1(x)e^{i\Ga x}dx.
\label{4.7}
\eeq
To derive the RHP, we first extend the Neumann boundary condition on the 
line $y=0$
\beq
\fr{\Md u}{\Md y}=f_-(x)+\Gvf_{2-}(x)+\Gvf_{2+}(x), \quad -\infty<x<\infty, \quad y=0^\pm,
\label{4.8}
\eeq
where
$f_-(x)=f(x)$ if $0<x<1$ and it vanishes otherwise, $\Gvf_\pm(x)$ are unknown functions
such that $\supp\Gf_{2-}\subset(-\infty,0]$ and $\supp\Gf_{2+}\subset[1,\infty)$.
In order to apply the Fourier transform to the boundary condition (\ref{4.8}),
we introduce the following integrals
$$
F^+(\Ga)=\int_0^1 f(x)e^{i\Ga x}dx, 
$$
\beq
\GF_2^-(\Ga)=\int_{-\infty}^0 \Gvf_{2-}(x)e^{i\Ga x}dx, \quad
\GF_2^+(\Ga)=\int_0^{\infty} \Gvf_{2+}(x+1)e^{i\Ga x}dx.
\label{4.9}
\eeq
We assert that  $\GF_1^+(\Ga)$ and $F^+(\Ga)$ are entire  functions
which are analytic in the upper half-plane and have an essential singularity
at the infinite point in the lower half-plane and also
\beq
\GF_1^+(\Ga)=e^{i\Ga}\GF_1^-(\Ga), \quad F^+(\Ga)=e^{i\Ga}F^-(\Ga),
\label{4.10}
\eeq
where
\beq
\GF_1^-(\Ga)=\int_{-1}^0 \Gvf_1(x+1)e^{i\Ga x}dx, \quad 
F^-(\Ga)=\int_{-1}^0 f(x+1)e^{i\Ga x}dx.
\label{4.11}
\eeq
This implies that the boundary condition (\ref{4.8}) is equivalent to the following
vector RHP:
\beq
\GF^+(\Ga)=G(\Ga)\GF^-(\Ga)+F^-(\Ga)J, \quad -\infty<\Ga<+\infty.
\label{4.12}
\eeq
Here,
$$
G(\Ga)=\left(
\begin{array}{cc}
e^{i\Ga}  & 0\\
-g(\Ga) & -e^{-i\Ga}\\
\end{array}
\right), \quad 
J=\left(
\begin{array}{c}
0\\
-1\\
\end{array}
\right),
$$
\beq
g(\Ga)=\Gg\sinh \Gg b_+\sinh \Gg b_- {\rm csch}\,\Gg(b_++b_-). 
\label{4.13}
\eeq

The new feature here is that the zeros and poles of the meromorphic entry of the matrix $G(\Ga)$,
the function $g(\Ga)$,
are not periodic, and we will pursue the RHP in a way different from the one
proposed in the previous section. First we factorize the meromorphic function $g(\Ga)$
\beq
g(\Ga)=\fr{K^+(\Ga)}{K^-(\Ga)},
\label{4.14}
\eeq
where
$$
K^\pm(\Ga)=(\Ga\pm k)^{\pm 1/2}K_0^\pm(\Ga), \quad -\infty<\Ga<+\infty,
$$
\beq
K_0(\Ga)=
\exp\left\{\fr{\Ga}{\pi i}
\int_0^\infty\log\fr{\sinh \Gg b_+ \sinh \Gg b_-}{\sinh \Gg(b_++b_-)}
\fr{d\Gb}{\Gb^2-\Ga^2}\right\}, \quad \Ga\notin(-\infty,+\infty),
\label{4.16}
\eeq
and $K_0^\pm(\Ga)$, $-\infty<\Ga<+\infty$, are defined  by the Sokhotsky-Plemelj formulas.
Then
we rewrite vector equation (\ref{4.12}) as
$$
K^-(\Ga)\GF_2^-(\Ga)-\Psi_-^-(\Ga)=-K^+(\Ga)\GF_1^+(\Ga)
-\fr{e^{i\Ga}K^+(\Ga)\GF_2^+(\Ga)}{g(\Ga)}
-\Psi_-^+(\Ga),
$$
\beq
\fr{\GF_2^+(\Ga)}{K^+(\Ga)}+\Psi_+^+(\Ga)=-\fr{\GF_1^-(\Ga)}{K^-(\Ga)}
-\fr{e^{-i\Ga}\GF_2^-(\Ga)}{K^-(\Ga) g(\Ga)}
+\Psi_+^-(\Ga),
\label{4.17}
\eeq
where
\beq
\Psi_-(\Ga)=\int_{-\infty}^\infty\fr{K^-(\Gb) F^+(\Gb)d\Gb}{\Gb-\Ga},\quad
\Psi_+(\Ga)=\int_{-\infty}^\infty\fr{F^-(\Gb)d\Gb}{K^+(\Gb) (\Gb-\Ga)}.
\label{4.17'}
\eeq
The right-hand sides of the first and second equations in (\ref{4.17})
are analytic everywhere in the domains $C^+: \I\Ga>0$ and $C^-:\I\Ga<0$, respectively,
except at the zeros of the function $g(\Ga)$.
Denote by $\Ga_{m\pm}$ the zeros of the functions $\sinh\Gg b_\pm$ lying
in the upper half-plane,
\beq
\Ga_{m\pm}=\sqrt{k^2-\fr{\pi^2 m^2}{b_\pm^2}}, \quad \I\Ga_{m\pm}>0, \quad m=1,2,\ldots.
\label{4.18}
\eeq
Then in $C^+$, the function $g(\Ga)$ has simple zeros at the points
$\Ga=k$ and $\Ga=\Ga_{m\pm}$, $m=1,2,\ldots$. In the lower half-plane, the 
zeros are $\Ga=-k$ and $\Ga=-\Ga_{m\pm}$, $m=1,2,\ldots$. 
In order to remove the simple poles of the right-hand sides of equations (\ref{4.17}),
we introduce the functions
$$
\GO^+(\Ga)=\fr{A_0^+}{\Ga+k}+\sum_{m=1}^\infty\left(
\fr{A_{m+}^+}{\Ga+\Ga_{m+}}+\fr{A_{m-}^+}{\Ga+\Ga_{m-}}
\right),
$$
\beq
\GO^-(\Ga)=\fr{A_0^-}{\Ga-k}+\sum_{m=1}^\infty\left(
\fr{A_{m+}^-}{\Ga-\Ga_{m+}}+\fr{A_{m-}^-}{\Ga-\Ga_{m-}}
\right).
\label{4.19}
\eeq
Assume that the residues of the functions $\GO^-(\Ga)$ and $\GO^+(\Ga)$ at their poles are chosen such that,
when they are subtracted from the left and right-hand sides of the first and second
equations in (\ref{4.17}), respectively, the poles are removed. On employing
the Liouville theorem we find the solution of the RHP
$$
\GF_2^-(\Ga)=\fr{\Psi_-^-(\Ga)+\GO^-(\Ga)}{K^-(\Ga)}, \quad
\GF_2^+(\Ga)=K^+(\Ga)[-\Psi_+^+(\Ga)+\GO^+(\Ga)], 
$$
$$
\GF_1^-(\Ga)=K^-(\Ga)[\Psi_+^-(\Ga)-\GO^+(\Ga)]-\fr{e^{-i\Ga}}{K^-(\Ga)g(\Ga)}[\Psi_-^+(\Ga)+\GO^-(\Ga)],
$$
\beq
\GF_1^+(\Ga)=-\fr{\Psi_-^+(\Ga)+\GO^-(\Ga)}{K^+(\Ga)}+\fr{e^{i\Ga}K^+(\Ga)}{g(\Ga)}[\Psi_+^+(\Ga)-\GO^+(\Ga)].
\label{4.20}
\eeq
It is directly verified that  $\GF_1^+(\Ga)=e^{i\Ga}\GF_1^-(\Ga)$.
The coefficients $A_0^\pm$, $A^\pm_{m+}$ and $A^\pm_{m-}$
need to be fixed by the conditions
\beq
 \mathop{\rm res}\limits_{\Ga=\pm k}\GF_1^\pm(\Ga)=0,
 \quad 
  \mathop{\rm res}\limits_{\Ga=\Ga_{m\pm}}\GF_1^+(\Ga)=0,
\quad 
  \mathop{\rm res}\limits_{\Ga=-\Ga_{m\pm}}\GF_1^-(\Ga)=0,
  \quad m=1,2,\ldots.
 \label{4.21}
\eeq
These conditions if satisfied guarantee that the zeros of the function $g(\Ga)$
are removable singularities of the functions $\GF_1^\pm(\Ga)$. On calculating the residues
we rewrite the conditions (\ref{4.21})
as a system of linear algebraic equations
$$
A_0^\mp\pm\Gd_0 e^{ik}\left[
\fr{A_0^\pm}{2k}+\sum_{m=1}^\infty\left( \fr{A_{m+}^\pm}{k+\Ga_{m+}}+
\fr{A_{m-}^\pm}{k+\Ga_{m-}}\right)
\right]=\pm\Gd_0 e^{ik}\Psi_\pm^\pm(\pm k),
$$
$$
A_{n\pm}^-+\Gd_{n\pm} e^{i\Ga_{n\pm}}\left[
\fr{A_0^+}{\Ga_{n\pm}+k}+\sum_{m=1}^\infty\left( \fr{A_{m+}^+}{\Ga_{n\pm}+\Ga_{m+}}+
\fr{A_{m-}^+}{\Ga_{n\pm}+\Ga_{m-}}\right)
\right]=\Gd_{n\pm} e^{i\Ga_{n\pm}}\Psi_+^+(\Ga_{n\pm}),
$$
$$
A_{n\pm}^+-\Gd_{n\pm} e^{i\Ga_{n\pm}}\left[
\fr{A_0^-}{\Ga_{n\pm}+k}+\sum_{m=1}^\infty\left( \fr{A_{m+}^-}{\Ga_{n\pm}+\Ga_{m+}}+
\fr{A_{m-}^-}{\Ga_{n\pm}+\Ga_{m-}}\right)
\right]
$$
\beq
=-\Gd_{n\pm} e^{i\Ga_ {n\pm}}\Psi_-^-(-\Ga_{n\pm}),
\quad n=1,2,\ldots.
\label{4.22}
\eeq
Here,
$$
\Gd_{n\pm}=\fr{(-1)^n(1+k/\Ga_{n\pm})[K_0^+(\Ga_{n\pm})]^2\sin[\pi n(b_++b_-)/b_\pm]}
{b_\pm\sin(\pi n b_\mp/b_\pm)},
$$
\beq
\Gd_0=\left(\fr{1}{b_+}+\fr{1}{b_-}\right)[K_0^+(k)]^2. 
\label{4.23}
\eeq
If we assume the symmetry of the problem, $b_-=b_+=b$, then the formulas can be simplified.
Indeed, $g(\Ga)$ now becomes $\fr12\Gg\tanh \Gg b$, and therefore its zeros
are given by $\Ga_m=\sqrt{k^2-\pi^2 m^2/b^2}$ $(\Ga_0=k)$, $\I\Ga_m>0$, $m=0,1,\ldots$.
The functions $\GO^\pm(\Ga)$ are represented by the series
\beq
\GO^\pm(\Ga)=\sum_{m=0}^\infty\fr{A_m^\pm}{\Ga\pm\Ga_m},
\label{4.24}
\eeq
and the infinite systems reduce to
\beq
A_n^\pm\mp\Gd_n e^{i\Ga_n}\sum_{m=0}^\infty\fr{A_m^\mp}{\Ga_n+\Ga_m}=h_n^\pm, \quad n=0,1,\ldots.
\label{4.25} 
\eeq
Here, 
\beq
h_n^\pm=\mp\Gd_ne^{i\Ga_n}\Psi_\mp^\mp(\mp\Ga_n), \quad 
\Gd_n=\fr{2(\Ga_n+k)}{\Ga_n b}[K_0^+(\Ga_n)]^2.
\label{4.26}
\eeq
Notice that in both cases, $b_-\ne b_+$ and $b_-=b_+$, the unknowns $A_n^+$
and $A_n^-$ exponentially decay as $n\to \infty$,
$|A_n^\pm|<c_\pm e^{qn}$, $c_\pm$ and $q$ are some positive constants.

\section{Conclusion}

We have developed  further the algorithm for the vector RHP when its matrix coefficient entries are meromorphic and almost periodic functions. In the simplest case, when the meromorphic functions have a finite number of poles and zeros (rational functions), regardless of the dimension of the vector RHP,
the exact solution can always been constructed. We have illustrated this approach by solving the inhomogeneous heat equation with a piece-wise constant diffusivity. In the case of two discontinuities, $0$ and $\infty$, we have derived a simple representation of the solution that generalizes  for  the discontinuous case the classical Poisson formula for an infinite rod.  

We have also shown that, when the meromorphic functions have periodic zeros and poles only, then it is possible to derive the Wiener-Hopf factors of the RHP matrix coefficient in  a closed form in terms of the hypergeometric functions. This technique has been employed for finding the exact solution of 
the vector RHP associated with 
a mixed boundary value problem for the Laplace equation in a wedge when on finite segments
of the wedge sides, the function is prescribed, and on two semi-infinite segments, the normal derivative is known, and no symmetry that would allow for decoupling of the problem is assumed.

In the general case, when the zeros or poles are not periodic, the solution can be derived by quadratures and some exponentially convergent series. The series coefficients solve an infinite system of linear algebraic equations of the second kind whose rate of converge is exponential. This technique has been used for solving
the Neumann boundary value problem for a strip with a finite slit inside parallel to the strip boundaries.

 \vspace{.1in}

 \vspace{.1in}
 
{\bf References}

\vspace{2mm}

\noindent 
Abrahams, I. D. \& Wickham, G. R. 1990 General Wiener-Hopf factorization of matrix kernels with exponential phase factors.  {\it SIAM J. Appl. Math.} {\bf 50}, 819-838. 

\noindent  Antipov, Y. A. 1987 Exact solution of the problem of pressing an annular stamp into a half-space. 
 {\it Dokl. Akad. Nauk Ukrain. SSR} Ser. A, no. 7, 29-33.
\noindent

\noindent  Antipov, Y. A. 1989 Analytic solution of mixed problems of mathematical physics
with  a change of boundary conditions over a ring. {\it Mechanics of solids} {\bf 24}, no. 3, 49-56.

\noindent  Antipov, Y. A. \& Arutyunyan, N. Kh. 1992 Contact problems in elasticity theory in the presence of friction and adhesion. {\it J. Appl. Math. Mech.}  {\bf 55}, 887-901.

\noindent  Antipov, Y.  A. 1995 The interface crack in elastic media in the presence of dry friction. 
 {\it J. Appl. Math. Mech.}  {\bf 59}, 273-287.

\noindent  Antipov, Y.  A. 2000 Galin's problem for a periodic system of stamps with friction and adhesion. 
{\it Internat. J. Solids Structures} {\bf 37}, no. 15, 2093-2125.

\noindent Antipov, Y.  A.  \& Schiavone, P. 2011 Integro-differential equation of a finite crack in a strip with surface effects. {\it Quart. J. Mech. Appl. Math.}  {\bf 64}, 87-106.

\noindent B\"ottcher,  A., Karlovich, Y. I. \&  Spitkovsky, I. M. 2002 {\it Convolution Operators and Factorization of Almost Periodic Matrix Functions}, Oper. Theory Adv. Appl., vol. 131, 
Basel: Birkh\"auser Verlag.

\noindent  Deconinck, B., Pelloni, B. \&
Sheils, N. E. 2014 Non-steady-state heat
conduction in composite walls. {\it Proc. R. Soc. A}  {\bf 464}: 20130605.

\noindent Fokas A. S. 2008 {\it A unified approach to boundary value problems}. CBMS-NSF Regional Conference Series in Applied Mathematics, vol. 78. Philadelphia, PA: Society for Industrial and Applied
Mathematics (SIAM).

\noindent   Ganin, M. P. 1963 On a Fredholm integral equation whose kernel depends on the difference of the arguments. {\it Izv. Vys\v{s}. U\v{c}ebn. Zaved. Matematika} no.2 (33), 31-43. 

\noindent   Karlovich, Y. I. \& Spitkovsky, I. M. 1983
On the Noethericity of some singular integral operators with matrix coefficients of class SAP and systems of convolution equations on a finite interval associated with them.  
{\it Dokl. Akad. Nauk SSSR} {\bf 269}, no. 3, 531-535. 

\noindent    Novokshenov, V. Y. 1980 Equations in convolutions on a finite interval and factorization of elliptic matrices. {\it Mat. Zametki}  {\bf 27}, 935-946.

\noindent Onishchuk, O. V. 1988 On a method of solving integral equations and its application to the problem of the bending of a plate with a cruciform inclusion. {\it J. App. Math. Mech.} {\bf 52}, 211-223.

\noindent Spitkovsky, I. M. 1989
Factorization of almost-periodic matrix functions. 
{\it Math. Notes}  {\bf 45}, 482-488.

\end{document}